\documentclass[preprintnumbers,superscriptaddress,amsmath,amssymb,prl,nofootinbib,twocolumn]{revtex4-2}
\usepackage{amssymb}
\usepackage{mathrsfs}
\usepackage{amsthm}
\usepackage{braket}
\usepackage{bm}
\usepackage{qcircuit}
\usepackage{graphicx}
\usepackage{amsfonts}
\usepackage[colorlinks,linkcolor=red,anchorcolor=blue,citecolor=green]{hyperref}
\usepackage{subfigure}
\usepackage{float}
\usepackage{enumerate}
\usepackage{multirow}

\usepackage{array}
\usepackage[figuresright]{rotating}
\usepackage{upgreek}

\begin{document}
	
\title{Geometric deformation and redshift structure \\ caused by plane gravitational waves}
	
\author{Ke Wang}
\affiliation{Division of Mathematical and Theoretical Physics, Shanghai Normal University, 100 Guilin Road, Shanghai 200234, P.R.China}
	
\author{Chao-Jun Feng}
\thanks{Corresponding author}
\email{fengcj@shnu.edu.cn}
\affiliation{Division of Mathematical and Theoretical Physics, Shanghai Normal University, 100 Guilin Road, Shanghai 200234,  P.R.China}
	
\begin{abstract}
The curved spacetime induced by gravitational waves can give rise to visual effects such as geometric distortions and redshift structures in the observed image. By establishing a mapping from the object's surface coordinates to the observer's screen coordinates, we study these effects in the context of plane gravitational waves. The simulation reveals that the image of an object doesn't merely seem compressed or stretched, but rather appears twisted and wobbled. Furthermore, the redshift structure on the object's surface appears to rotate as a whole. This outcome offers an intuitive depiction of the lensing effect in plane gravitational wave spacetimes.
\end{abstract}

\maketitle

\section{Introduction}

Similar to the lensing phenomenon caused by light rays bending around massive objects, fluctuations in spacetime caused by gravitational waves can also cause light rays to bend, resulting in lensing.
Unlike gravitational lensing in an ideal static spacetime, gravitational wave spacetime is dynamic, and the lensing phenomenon produced by gravitational waves should also be dynamic.
Since the strain amplitude of gravitational waves tends to be extremely small, it generally appears as spacetime perturbation terms in theoretical studies\cite{abbott_observation_2016}. Therefore, researchers focus more on the perturbation effects of gravitational waves on ordinary lens systems \cite{Allen:1989my,Frieman:1994pe,Pyne:1995iy,Bar-Kana:1996dkw,Book:2010pf,Wang:2019skw}.
Although these calculations take into account the effects of gravitational waves, they are dominated by the primordial lens system and therefore neglect many details of the effects of gravitational waves. 
A preliminary simulation of the visual effect of gravitational wave spacetimes will be conducted to show an intuitive picture of gravitational wave lensing effects. Generally speaking, when describing the local effect far away from the source of gravitational waves, the precise generation mechanism of gravitational waves is not important, and the spacetime in the local region far away from the source is approximated to a plane gravitational wave spacetime\cite{Bondi:1957dt,Blau:2006ar}. 
Although the plane gravitational wave is highly idealized, it is a useful model for studying the lensing effect in non-stationary spacetimes. As an exact solution of the Einstein equation in vacuum, the simple mathematical representation of the plane gravitational wave allows us to study gravitational lensing accurately without any weak field or Newtonian approximation. 
As early as 1965, Penrose noticed the extraordinary causal structure of plane wave spacetime and found the focusing effect of past light cones, called ``astigmatic focusing" \cite{penrose1965remarkable},  which led to multiple imaging in plane wave spacetimes. The geodesic and causal structure of plane wave spacetime have also been investigated in detail in later years \cite{kovner1990fermat,faraoni1992nonstationary,ehrlich1992gravitational,hubeny2003causal}.
Although these early studies focused on the geometric properties of spacetime and various mathematical principles, they were closely related to gravitational wave lensing effects and have guided some recent discussions on gravitational wave lensing \cite{Vollick:1991jf,Damour:1998jm,faraoni1998multiple,Harte:2012jg,Harte:2015ila}.  These studies revealed the rich lensing phenomenon in plane gravitational wave spacetimes.

This paper is organized as follows: In the next section, a general method of imaging in curved spacetimes is introduced. Following that, the subsequent two sections detail the calculation and simulation of the geometric distortion and redshift structure caused by a linear plane gravitational wave. Finally, in the last section, conclusions and discussions will be provided.

\section{General setup}\label{sec:gs}
To perceive an object within curved spacetime, one must meticulously trace each emitted ray of light from the object's surface to the observer's position. Typically, establishing a correspondence between the observer's screen coordinates \(\bm{\theta}\) and the object's surface coordinates \(\bm{x}\) is essential, which can be expressed through the mapping:
\begin{align}
	\label{eq:mapping}
	\mathcal{F}:\bm{\theta }\to\bm{x} \,.
\end{align}
In a flat spacetime, light travels in a straight line, rendering the mapping \(\mathcal{F}\) a simple projection transformation. However, within curved spacetime, the trajectory of light propagation and its measurement necessitate intricate processing. Each photon reaching the observer is projected onto a point on their screen based on its locally measured four-momentum. The mapping relationship \(\mathcal{F}\) is typically derived through two steps: initially, determining the initial conditions of the associated null geodesic for each point \(\bm{\theta }\) on the screen, and subsequently retracing the light's path back to a point \(\bm{x}\) on the surface by solving the geodesic equation. After obtaining the mapping, the geometric deformation and redshift structure of the object image can be obtained. \\

\subsection{Step 1: Initial conditions and local measurement}
\label{2.1}
As is well known, the geodesic equation in general relativity comprises four second-order differential equations. Solving these equations necessitates eight integral constants, which are associated with the initial position and four-momentum of an object. In the scenario under consideration, spacetime exhibits certain symmetries, enabling us to perform the first integration of the photon's geodesic equation. It is established that the geodesic equation for photons satisfies \(d\tau^2=0\), indicating three constants in the photon's four-momentum. However, the observer's screen is merely a two-dimensional plane with two coordinates, rendering it impossible to fully determine the photon's four-momentum. Complete determination requires knowledge of the photon's energy. Nonetheless, the photon's energy does not influence its trajectory. Therefore, when solely focused on the photon's trajectory, the degrees of freedom associated with the photon's energy can be eliminated by redefining the affine parameter \(\sigma\) as 
\begin{align}
	\label{eq:2.1-1}
	\sigma =f\left(x^{\mu }\right) \,,
\end{align}
where \(f\) is an arbitrary function of the coordinates \(x^{\mu }\). Utilizing appropriate affine parameters can significantly simplify the geodesic equation. For example, in Kerr black hole spacetimes in Boyer-Lindquist coordinates, employing "Mino time" \cite{mino_perturbative_2003}  as the null geodesic affine parameter can decouple the equations of \(r\) and \(t\). However, it's crucial to note that certain choices of affine parameters may result in the invalidation of the geodesic equation.

Given the coordinates \(\bm{\theta}\) of photons reaching the observer screen, a connection can be established between them and the four-momentum \(p^{\hat\mu }\), denoting the measurements made by the local observer, as follows:
\begin{align}
	\label{eq:2.1-2}
	\bm{\theta }=\mathcal{P}\left(p^{\hat{\mu }}\right) \,,
\end{align}
where the specific form of the function \(\mathcal{P}\) can be determined by the observer's screen model. The local measurement of the photon's four-momentum \(p^{\hat{\mu }}\) is obtained by projecting it onto the observer's local tetrad \(\left\{\bm{e}_{\hat{\mu}}\right\}\), expressed as:
\begin{align}
	\label{eq:2.1-3}
	p^{\hat{\mu }}=\eta ^{\hat{\mu }\hat{\nu }}\bm{e}_{\hat{\nu }}^{\alpha }p_{\alpha } \,,
\end{align}
where \(\bm{e}_{\hat{\nu }}^{\alpha }=\bm{e}_{\hat{\nu }}\cdot \partial _{\alpha }\) and \(\bm{e}_{\hat{\mu }}\cdot \bm{e}_{\hat{\nu }}=\eta_{\hat{\mu }\hat{\nu }}\).

\subsection{Step 2: Back ray-tracing}
\label{2.2}
When tackling the second-order geodesic equation:
\begin{align}
	\frac{d^2 x^{\mu }}{d \sigma ^2}=g^{\mu \nu }\left(\partial _{\nu }\mathcal{L}-\frac{d g_{\nu \beta }}{d \sigma }\frac{d x^{\beta }}{d \sigma }\right) \,,
\end{align}
symmetry is typically leveraged as the initial step to integrate it for the first time and obtain the first-order equation. However, the specific form of the metric needs to be provided beforehand. Alternatively, the Hamiltonian equation can also directly provide the specific form of these equations. Solving geodesic equations often involves intricate technicalities, and researchers have developed numerous analytical and numerical methods for solving geodesic equations across various spacetimes \cite{felice_effects_1979,gralla_null_2020,dexter_fast_2009}.

\section{Gemetric deformation}
\label{3}
\begin{figure}[htbp]
	\centering
	\includegraphics[width=0.4\textwidth]{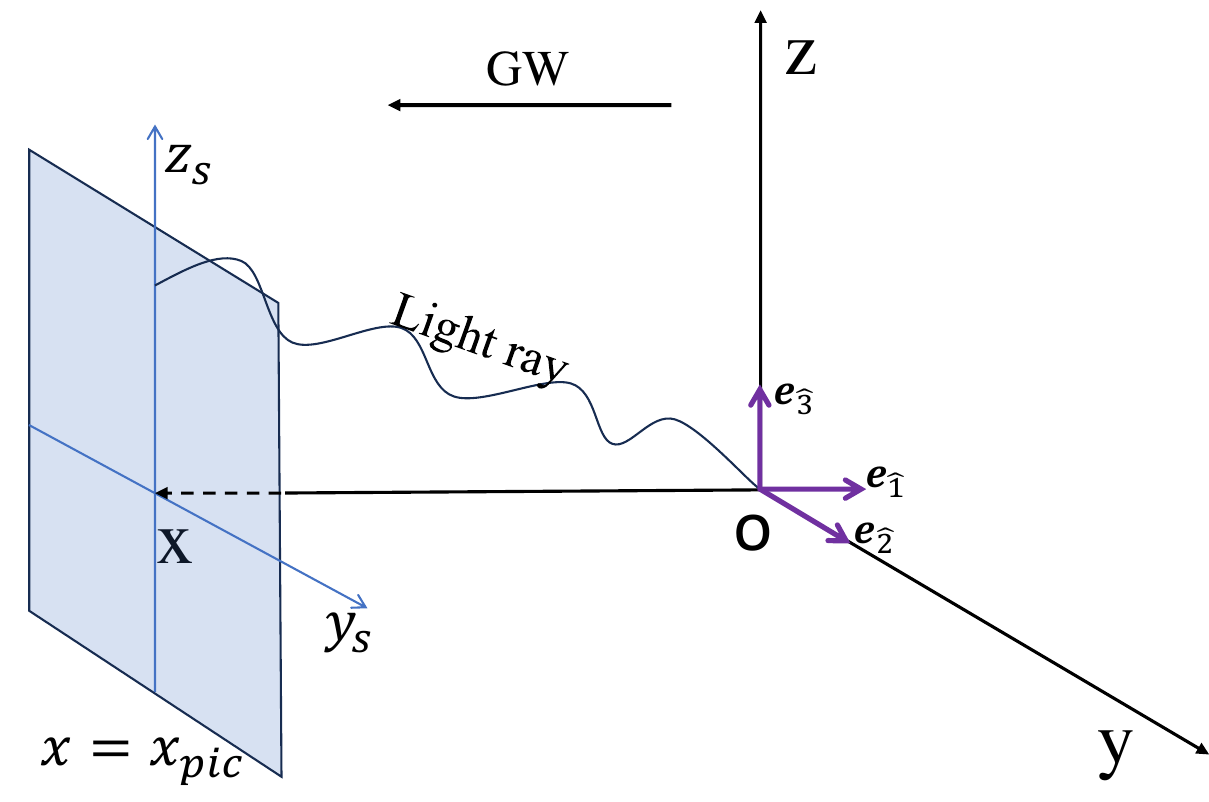}
	\caption{The three-dimensional coordinate system of a flat spacetime affected by a plane gravitational wave propagating along the x-axis.  The stationary observer is situated at the origin and gazes at a vertically positioned picture located at \(x=x_\text{pic}\).}
	\label{fig:model1}
\end{figure}

Consider a flat spacetime traversed by a linear plane gravitational wave in Cartesian coordinates \((t,x,y,z)\). The gravitational wave propagates along the x-axis direction under the transverse-traceless (TT) gauge condition. The observer is positioned at the coordinate origin, directing their gaze towards an image located at \(x_{\text{pic}}\), as illustrated in \autoref{fig:model1}.
To streamline the calculation and enhance clarity, we initially perform a coordinate transformation, defining \(u=(t-x)/\sqrt{2}\) and \(v=(-t-x)/\sqrt{2}\), where \(u\) and \(v\) denote the retarded time and the advanced time, respectively. Subsequently, the metric $ds^2=g_{\mu\nu}dx^\mu dx^\nu$ is expressed as:
\begin{eqnarray}
	ds^2= 2d u dv+(1-h_+)dy^2+(1+h_+)dz^2 -2h_{\times}dy dz\,.
\end{eqnarray}
where \(h_{+}\) and \(h_{\times}\) represent the two modes of the gravitational wave, which are solely functions of \(u\).

The Lagrangian describing the motion of the object is given by $\mathcal{L}=\frac{1}{2}g_{\mu \nu }\dot{x}^{\nu }\dot{x}^{\mu }$ or, explicitly, as follows:
\begin{eqnarray}
		\mathcal{L}= \dot{u}\dot{v}+\frac{\left(1-h_+\right)}{2}\dot{y}^2+\frac{\left(1+h_+\right)}{2}\dot{z}^2-h_{\times}\dot{y}\dot{z}\,,\label{eq:lag}
\end{eqnarray}
where the overdot denotes the derivative with respect to the affine parameter.
It's evident that the Lagrangian \eqref{eq:lag} is independent of the coordinates \(v\), \(y\), and \(z\). Consequently, according to the Euler-Lagrange equation, \(\partial \mathcal{L}/\partial \dot v\), \(\partial \mathcal{L}/\partial \dot y\), and \(\partial \mathcal{L}/\partial \dot z\) are conserved quantities. These three conserved quantities are denoted respectively by \(E\), \(\alpha\), and \(\beta\) as follows:
\begin{align}
	E &= p_v=g_{1\nu }\dot{x}^{\nu }=\dot{u}    \,,\\
	\alpha &=p_y=g_{2\nu }\dot{x}^{\nu }=\left(1-h_+\right)\dot{y}-h_{\times}\dot{z}   \,,\\
	\beta &=p_z=g_{3\nu }\dot{x}^{\nu }=\left(1+h_+\right)\dot{z}-h_{\times}\dot{y} \,.
\end{align}

The Hamiltonian gives the fourth constant $\mu$
\begin{eqnarray}
	\nonumber
		\mathcal{H}&=&\frac{1}{2}g^{\mu \nu }p_{\mu }p_{\nu }=\frac{\mu }{2} \\
	&=&p_vp_u+\frac{1+h_+}{2\Delta}p_y^2+\frac{1-h_+}{2\Delta}p_z^2+\frac{h_x}{\Delta}p_yp_z\,,
\end{eqnarray}
where
\begin{eqnarray}
	\Delta =  1-h_+^2-h_\times^2\,.
\end{eqnarray}
Actually, the geodesic equation can be given by the Hamiltonian equation by  $\dot x^\mu = \partial \mathcal{H}/\partial  p_\mu$
\begin{eqnarray}
			\dot{u}&=&E   \,,\label{eq:geodesic1} \\
	\dot{v}&=&\frac{\mu}{2E} - \frac{ \Gamma}{2E\Delta}  \,,\label{eq:geodesic2}\\
	\dot{y}
	&=&\frac{\left(1+h_+\right)\alpha +h_\times\beta }{\Delta}  \,,\label{eq:geodesic3}\\
	\dot{z}
	&=&\frac{\left(1-h_+\right)\beta +h_\times\alpha }{\Delta} \,,\label{eq:geodesic4}
\end{eqnarray}
where
\begin{eqnarray}
	\Gamma =  \left(1+h_+\right)\alpha^2+\left(1-h_+\right)\beta ^2+2h_\times\alpha \beta \,.
\end{eqnarray}
When we take $\sigma=u$ in \eqref{eq:2.1-1}, the constant $E$ in equation \eqref{eq:geodesic1} becomes one. Furthermore, since the geodesic of photons satisfies $\mu=0$, one can obtain the conjucated momentum $p_\mu = g_{\mu\nu}p^\mu =  g_{\mu\nu}\dot x^\mu$ as the following
\begin{equation}
	\label{eq:momentum1}
	p_{\mu }=\left(-\frac{\Gamma }{2\Delta},1,\alpha ,\beta \right) \,.
\end{equation}

The observer's local orthonormal frame \(\left\{\bm{e}_{\hat{\mu }}\right\}\) is established as \cite{Bini:2000xj}
\begin{eqnarray}
	\bm{e}_{\hat{0}}&=&\frac{1}{\sqrt{2}}\partial _u-\frac{1}{\sqrt{2}}\partial _v  \,,\\\label{eq:tetrad1}
	\bm{e}_{\hat{1}}&=&\frac{1}{\sqrt{2}}\partial _u+\frac{1}{\sqrt{2}}\partial _v  \,,\\\label{eq:tetrad2}
	\bm{e}_{\hat{2}}&=&\frac{1}{\sqrt{1-h_+}}\partial _y   \,,\\\label{eq:tetrad3}
	\bm{e}_{\hat{3}}&=&\Delta^{-1/2}\left[\frac{h_\times}{\sqrt{1-h_+}}\partial _y+\sqrt{1-h_+}\partial _z\right] \,. \label{eq:tetrad4}
\end{eqnarray}
Substituting the photon's four-momentum \eqref{eq:momentum1} and the observer's frame \eqref{eq:tetrad1} into \eqref{eq:2.1-3}, we get
\begin{eqnarray}
		p^{\hat{t}}
		&=&\frac{1}{\sqrt{2}}+
	\frac{\Gamma }{2\sqrt{2}\Delta}  \,,\\\label{eq:localmomentum1}
	p^{\hat{x}}
	&=&\frac{1}{\sqrt{2}}
	-\frac{\Gamma }{2\sqrt{2}\Delta}  \,,\\\label{eq:localmomentum2}
	p^{\hat{y}}
	&=& \frac{\alpha}{\sqrt{1-h_+}}  \,,\\\label{eq:localmomentum3}
	p^{\hat{z}}
	&=& \Delta^{-1/2}\left[\frac{h_\times}{\sqrt{1-h_+}}\alpha+\sqrt{1-h_+}\beta\right] \,.\label{eq:localmomentum4}
\end{eqnarray}

In our scenario, the projection relationship \eqref{eq:2.1-2} between the screen coordinates \((y_s,z_s)\) and the local measured four-momentum is
\begin{align}
	\label{eq:projection}
	y_s=-x_\text{pic}\frac{p^{\hat{y}}}{p^{\hat{x}}} \,, \qquad  z_s=-x_\text{pic}\frac{p^{\hat{z}}}{p^{\hat{x}}} \,.
\end{align}
Combining \eqref{eq:localmomentum1}- \eqref{eq:localmomentum4} and \eqref{eq:projection}, the above relation becomes
\begin{eqnarray}
	\frac{y_s}{x_{\text{pic}}}&=&-\frac{2 \sqrt{2} \alpha  \Delta }{\sqrt{1-h_+}(2\Delta - \Gamma)}\,,  \label{eq:ys} \\
	\frac{z_s}{x_{\text{pic}}}&=&-\frac{2 \sqrt{2} \sqrt{\Delta}}{\sqrt{1-h_+} (2\Delta-\Gamma) }  \bigg[ h_\times  \alpha +(1-h_+)\beta\bigg]\,.\label{eq:zs}
\end{eqnarray}
For any screen coordinates \((y_s,z_s)\), the constants \(\alpha\) and \(\beta\) in the four-momentum \eqref{eq:momentum1} of the corresponding photon are obtained by inversely solving \eqref{eq:ys} and \eqref{eq:zs}. Then, the complete geodesic trajectory is obtained by integrating the geodesic equations \eqref{eq:geodesic1}-\eqref{eq:geodesic4}, resulting in:
\begin{eqnarray}
		x(\sigma_s)&=&-\int_{\sigma_s}^{\sigma_O} \frac{1}{\sqrt{2}}
	\left(1-\frac{\Gamma}{2\Delta} \right) d \sigma +x_O      \,,\label{eq:trajectory1}\\
	y(\sigma_s)&=&-\int_{\sigma_s}^{\sigma_O} \frac{\left(1+h_+\right)\alpha +h_\times\beta }{\Delta} d \sigma +y_O     \,, \label{eq:trajectory2}\\
	z(\sigma_s)&=&-\int_{\sigma_s}^{\sigma_O} \frac{\left(1-h_+\right)\beta +h_\times\alpha }{\Delta} d \sigma +z_O    \,.\label{eq:trajectory3}
\end{eqnarray}
where \((x_O,y_O,z_O)\) denote the position of the observer, \(\sigma_s\) and \(\sigma_O\) denote the initial and final affine parameters, respectively. Utilizing the geodesic trajectory equations \eqref{eq:trajectory1}, we trace the photon back to the location of the source image \(x=x_\text{pic}\). The initial affine parameter \(\sigma_s\) of the photon is solved from \(x(\sigma_s)=x_\text{pic}\). Finally, the initial position of the photon is \((x_\text{pic},y(\sigma_s),z(\sigma_s))\).

At this point, we have established a mapping \(\mathcal{F}\) from the screen coordinates \((y_s,z_s)\) to the surface coordinates \((y,z)\) of the object.
The specific form of the mapping \(\mathcal{F}\) can be expressed as
\begin{eqnarray}
	\nonumber
	\{y,z\}&=&\mathcal{F}(y_s,z_s)\\
	\nonumber
	&=&\bigg\{y[\sigma_s;\alpha(y_s,z_s),\beta(y_s,z_s)],\\
	&& z[\sigma_s;\alpha(y_s,z_s),\beta(y_s,z_s)]\bigg\} \,.	\label{eq:mathcalF}
\end{eqnarray}

\begin{figure*}[htbp]
		\centering
		\includegraphics[width=\textwidth]{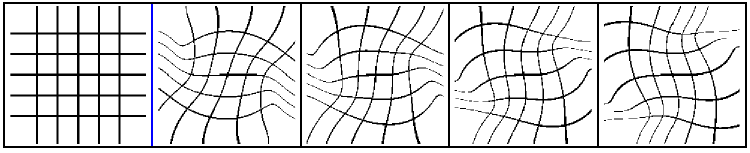}
		
		\caption{The left image is the original image, and the four images on the right are the visual effects caused by gravitational waves. The phase of the gravitational waves gradually increases from right to left.}
		
		\label{fig:grid}
	\end{figure*}

Consider a simple gravitational wave waveform given by:
\begin{equation}
	h_+(\sigma)=\frac{1}{3}\sin\sigma   \,,\quad 
	h_\times(\sigma)=\frac{1}{3}\cos\sigma  \,,
\end{equation}
and set the distance of the image from the observer to \(x_\text{pic}=10\). We initialize the image as a grid style. Subsequently, we utilize \eqref{eq:mathcalF} to generate visual renderings at several moments.
Figure \ref{fig:grid} illustrates the geometric distortion of mapping \(\mathcal{F}\).

In modern textbooks on gravitational waves \cite{ferrari2020general}, the effect of gravitational waves is often depicted as a ring being stretched or compressed into an ellipse. However, this representation is not an accurate depiction of the real visual phenomenon.
Our simulations illustrate the intricate geometric distortions induced by linear plane gravitational waves. As depicted in Figure \ref{fig:grid}, objects do not merely appear compressed or stretched as the gravitational wave passes through. Instead, they exhibit distortion and a wobbling effect.

\section{Redshift structure}
\begin{figure*}[htbp]
	\centering
	\includegraphics[width=\textwidth]{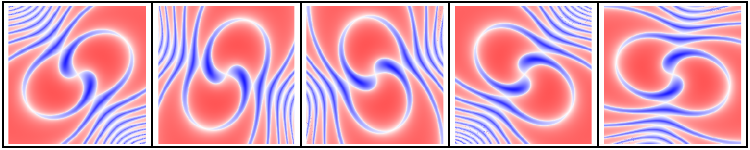}
	
	\caption{A map of the red-blue shift caused by gravitational waves, with progressively increasing phases of gravitational waves from right to left}
	
	\label{fig:redshift}
\end{figure*}

Gravitational waves not only induce geometric distortions in vision but also induce changes in the frequency of received photons. When the frequency of light received by the observer is greater or less than the frequency of light emitted by the object, the object will exhibit a blueshift or redshift feature. We can quantify the degree of change in the photon's frequency with the redshift factor
\begin{eqnarray}
		g=\frac{\nu _{\text{obs}}}{\nu _{\text{em}}}=\frac{k_{\mu }u_{\text{obs}}^{\mu }}{k_{\mu }u_{\text{em}}^{\mu }}
	=\frac{ 2+\Gamma/\Delta  \bigg|_{\sigma_O}}
	{2+\Gamma/\Delta  \bigg|_{\sigma_s}}  \,.
\end{eqnarray}
To visually illustrate the redshift structure of the image, we compute the redshift coefficients for each point on the observer screen and assign different shades of red or blue depending on the magnitude of the redshift coefficients. Results are depicted in Figure \ref{fig:redshift}.

In contrast to the situation in black hole lensing images, the redshift near the event horizon increases rapidly towards infinity. This infinite redshift boundary delineates the outline of the dark region at the center of the black hole image \cite{Wang:2023fge}\cite{qu_images_2023}.
However, unlike black hole lensing images, there are no extreme redshift features in lensed images in linear plane gravitational wave spacetime. In our simulation, the redshift magnitude of each point in Figure~\ref{fig:redshift} changes over time. The blueshift regions are interconnected in a spiral structure, and the overall redshift structure of the image exhibits a rotating phenomenon during the passage of the gravitational wave.

\section{Conclusions and Discussions}
\label{4}

In a curved spacetime background, the process of perceiving and visualizing an object can be understood as a mapping from the object's surface coordinates to the observer's screen coordinates. Using the curved spacetime generated by gravitational waves as an example, we calculate and  simulate the visual phenomena of the image over time. This includes observing the geometric distortion and the redshift structure induced by plane gravitational waves. In our current simulation, we have not considered the time delay effects \cite{wambsganss1998gravitational} of each point and the influence of nonlinear gravitational waves. These aspects remain avenues for future research.

For real astrophysical observations, factors such as multi-directional waves, non-radiative metric perturbations, and other effects may significantly influence the observed system. However, despite these complexities, results obtained using plane wave descriptions can still be valuable. They can aid in constructing specific hypotheses, the generality of which can later be tested using more intricate models. Moreover, the technical insights gained from plane wave descriptions can potentially inspire simplifications in more general calculations. Thus, while plane wave descriptions may not capture all aspects of the observed phenomena, they serve as a useful starting point in astrophysical investigations.

Furthermore, in early studies on the structure of past light cones in nonlinear plane wave spacetimes \cite{hubeny2003causal,Candela:2002rr}, the phenomenon of multiple imaging due to the observer's light cone was mentioned. However, the current discussion of multiple imaging exists primarily in mathematical analysis \cite{faraoni1998multiple,Petters:2010an,Harte:2012jg}. Demonstrating this phenomenon through simulation in specific models is challenging. Additionally, it remains unclear whether gravitational waves capable of inducing multiple imaging can be generated in the universe.
	
\section{Acknowledgments}
This work is supported by National Science Foundation of China grant No. 11105091.
	

\bibliographystyle{unsrt}
\bibliography{ref}
\end{document}